\documentclass[conference]{IEEEtran}
\IEEEoverridecommandlockouts
\usepackage{cite}
\usepackage{amsmath,amssymb,amsfonts}
\usepackage{algorithmic}
\usepackage{graphicx}
\usepackage{textcomp}
\usepackage{xcolor}
\usepackage{booktabs}
\usepackage{graphicx}
\usepackage[a-2b,mathxmp]{pdfx}
\def\BibTeX{{\rm B\kern-.05em{\sc i\kern-.025em b}\kern-.08em
    T\kern-.1667em\lower.7ex\hbox{E}\kern-.125emX}}
\usepackage{booktabs}
\begin{document}
\title{SCALAR: A Part-of-speech Tagger for Identifiers}

\author{
\IEEEauthorblockN{
    Christian D. Newman\IEEEauthorrefmark{1}, 
    Brandon Scholten\IEEEauthorrefmark{2}, 
    Sophia Testa\IEEEauthorrefmark{2}, 
    Joshua A. C. Behler\IEEEauthorrefmark{2},\\ 
    Syreen Banabilah\IEEEauthorrefmark{2}, 
    Michael L. Collard\IEEEauthorrefmark{3}, 
    Michael J. Decker\IEEEauthorrefmark{4}, 
    Mohamed Wiem Mkaouer\IEEEauthorrefmark{5},\\
    Marcos Zampieri\IEEEauthorrefmark{6}, 
    Eman Abdullah AlOmar\IEEEauthorrefmark{7}
    Reem Alsuhaibani\IEEEauthorrefmark{8}
    Anthony Peruma \IEEEauthorrefmark{9}\\
    Jonathan I. Maletic\IEEEauthorrefmark{2}
}

\IEEEauthorblockA{\IEEEauthorrefmark{1}\textit{Department of Software Engineering}, 
\textit{Rochester Institute of Technology},
Rochester, NY, USA,
0000-0002-8838-4074}

\IEEEauthorblockA{\IEEEauthorrefmark{2}\textit{Department of Computer Science},
\textit{Kent State University},
Kent, OH, USA,
\\(bscholte, stesta4, jbehler1, sbanabil, jmaletic)@kent.edu}

\IEEEauthorblockA{\IEEEauthorrefmark{3}\textit{Department of Computer Science},
\textit{University of Akron},
Akron, OH, USA,
collard@akron.edu}
\IEEEauthorblockA{\IEEEauthorrefmark{4}\textit{Department of Software Engineering},
\textit{Bowling Green State University},
Bowling Green, OH, USA,
mdecke@bgsu.edu}

\IEEEauthorblockA{\IEEEauthorrefmark{5}\textit{Department of Computer Science},
\textit{University of Michigan Flint},
Flint, MI, USA,
mmkaouer@umich.edu}

\IEEEauthorblockA{\IEEEauthorrefmark{6}George Mason University 
\textit{School of Computing},
Fairfax, VA, USA,
mzampier@gmu.edu}

\IEEEauthorblockA{\IEEEauthorrefmark{7}\textit{Stevens Institute of Technology},
\textit{School of Engineering and Science},
Hoboken, NJ, USA,
ealomar@stevens.edu}

\IEEEauthorblockA{\IEEEauthorrefmark{8}\textit{Prince Sultan University} 
\textit{Department of Software Engineering},
Riyadh, Saudi Arabia,
rsuhaibani@psu.edu.sa}

\IEEEauthorblockA{\IEEEauthorrefmark{9}\textit{University of Hawaii at Manoa} 
\textit{Department of Information and Computer Sciences},
Honolulu, HI, USA,
peruma@hawaii.edu}
}

\maketitle

\begin{abstract}
The paper presents the Source Code Analysis and Lexical Annotation Runtime (SCALAR), a tool specialized for mapping (annotating) source code identifier names to their corresponding part-of-speech tag sequence (grammar pattern). SCALAR's internal model is trained using scikit-learn's GradientBoostingClassifier in conjunction with a manually-curated oracle of identifier names and their grammar patterns. This specializes the tagger to recognize the unique structure of the natural language used by developers to create all types of identifiers (e.g., function names, variable names etc.).  SCALAR's output is compared with a previous version of the tagger, as well as a modern off-the-shelf part-of-speech tagger to show how it improves upon other taggers' output for annotating identifiers. The code is available on Github\footnote{https://github.com/SCANL/scanl\_tagger}
\end{abstract}


\section{Introduction}
The identifiers developers create represent a significant amount of the information other developers must use to understand related code. Given that identifiers represent, on average, 70\% of the characters in a code base\cite{Deissenbock:2005}, and developers spend more time reading code than writing \cite{Corbi1989,Martin:2008}, it is important for researchers to better understand of how identifiers convey information, and how they can be improved to increase developer reading efficiency. This problem is complicated by the fact that there are multiple comprehension styles \cite{vonMayrhauser:1997,Fisher:2006}, and the influence of identifier naming at varying levels of experience is currently understudied, especially in education contexts \cite{Felienne:2024, Glassman2015FoobazVN}. Further, while there are many studies that correlate the influence of identifiers (and various related characteristics) on comprehension \cite{Schankin:2018,Hofmeister:2017,butler2010exploring,feitelson:2017}, research has not found formal rules or unification among the outcomes of these studies due to the complexity of the problem, and the need for further research.

Thus, research is needed to improve identifier naming practices. Many approaches try to improve identifiers through predicting words that should appear in a name \cite{Allamanis:2015, Liu:2019} or analyzing/normalizing identifiers \cite{Jingxuan:2023, naser:2023, Arnaoudova:2014, Host:2009, Binkley:2011, Gupta:2013, shinpei:2021, shinpei:2022, newmanabbrev, Hill:2014} to understand how well they fit within a coding context, or make them easier to process. However, a significant challenge in identifier naming research lies in measuring the semantics of identifier names, then using that information to critique/generate better names. 

Measurement of identifier name semantics requires mapping between the terms and their meaning as an identifier (i.e., a sequence of terms). The ability to do this in a formal way allows us to recommend sequences of terms based on an understanding of the semantics they typically convey together. It also allows us to determine when the terms used in a sequence are inappropriate, since certain sequences are uncommon or even known to be anti-patterns. This is a challenge, as there are many ways to cluster identifiers using the terms used to construct them. Prior research shows that using grammar patterns to cluster/group identifiers with similar part-of-speech (PoS) sequences is an effective way to study how different types of identifiers convey meaning \cite{Newman_GP, newman2021Catalogue, peruma_test_name_gp}, and has the advantage that every term that is part of a natural language can be associated with a PoS tag. In addition, PoS tags are a known and well-supported means to formally define the function of a term within a sequence. Thus, we can group identifiers by grammar patterns (i.e., their PoS sequence) to measure the information they convey, and what patterns are most commonly used to convey different types of information. This is in contrast, and potentially complementary, to other approaches that cluster raw terms, or use vector representations.

Thus, we present SCALAR: A part-of-speech tagging approach specialized for source code identifiers. SCALAR is explicitly designed to support the generation of grammar pattern sequences to support future research and development of techniques that leverage grammar patterns. The goal of this paper is to show the effectiveness of SCALAR for generating grammar patterns.  SCALAR can be used in the future to improve techniques to analyze, recommend, and critique identifiers, by generating grammar patterns that can be used to understand identifier name meanings\cite{Newman_GP, newman2021Catalogue, peruma_digits, peruma_test_name_gp}. SCALAR is still very much in development as we apply it in our research and modify it to assist. However, it is a working tool that will be useful for others in identifier-oriented work.

\begin{table}[]
\centering
\caption{Part-of-speech categories in dataset and supported by SCALAR}
\label{tab:posusedtable}
\resizebox{\columnwidth}{!}{%
\begin{tabular}{@{}ccc@{}}
\toprule
\textbf{Abbreviation} & \textbf{Expanded Form} & \textbf{Examples}                    \\ \midrule
N                     & noun                   & stack, function, language            \\ \midrule
DT                    & determiner             & the, this, that, these, those, which \\ \midrule
CJ                    & conjunction            & and, for, nor, but, or, yet, so      \\ \midrule
P &
  preposition &
  \begin{tabular}[c]{@{}c@{}}behind, in front of, at, under,\\ beside, above, beneath, despite\end{tabular} \\ \midrule
NPL                   & noun plural            & strings, identifiers, classes        \\ \midrule
NM &
  \begin{tabular}[c]{@{}c@{}}noun modifier \\ (\textbf{noun adjunct}, adjective)\end{tabular} &
  \begin{tabular}[c]{@{}c@{}}\textbf{employee}Name, \textbf{token}Parser,\\ dynamic\end{tabular} \\ \midrule
V                     & verb                   & run, execute, implement, develop     \\ \midrule
VM                    & verb modifier (adverb) & quickly, safely, eventually          \\ \midrule
PR &
  pronoun &
  \begin{tabular}[c]{@{}c@{}}she, he, her, him, it, we,\\ us, they, them, I, me, you\end{tabular} \\ \midrule
D                     & digit                  & 1, 2, 10, 4.12, 0xAF                 \\ \midrule
PRE                   & preamble*              & Gimp, GLEW, GL, G                    \\ \bottomrule
\end{tabular}%
}
\end{table}
\section{Related Work}
POSSE \cite{Gupta:2013} and SWUM \cite{HillSWUM:2010}, and SCANL tagger \cite{newman_ensemble} are part-of-speech taggers created specifically to be run on software identifiers; they are trained to deal with the specialized context in which identifiers appear. Both POSSE and SWUM take advantage of static analysis to provide annotations. For example, they will look at the return type of a function to determine whether the word \textit{set} is a noun or a verb. Additionally, they are both aware of common naming structures in identifier names. For example, methods are more likely to contain a verb in certain positions within their name (e.g., at the beginning) \cite{Gupta:2013,HillSWUM:2010}. They leverage this information to help determine what POS to assign different words. Olney et al. \cite{Olney2016} compared taggers for accuracy on 200+ identifiers, but only on Java method names. They found that SWUM and POSSE were the most accurate taggers for source code at the time of publication. Newman et al. \cite{Newman_GP} compared the same taggers but on a larger dataset (1,335 identifiers) and five identifier categories: function, class, attribute, parameter, and declaration statement. They found that SWUM was the most accurate overall, with an average accuracy around 59.4\% at the identifier level. Later, Newman et al. created a new tagger and compared with SWUM, POSSE, and Stanford\cite{Toutanova:StanfordTagger}, finding that their new tagger exceeded the others' performance metrics on identifier names \cite{newman_ensemble}.

\section{Methodology}
The core of SCALAR is a GradientBoostingClassifier that is trained using the combination of two manually-curated data sets of identifiers and their corresponding grammar patterns. The first data set is called the General Grammar Dataset (GGD). The second dataset is called the Closed Grammar Dataset (CGD). The first dataset is used to train the first iteration of SCALAR \cite{newman_ensemble}, while the second is created to improve on the original: It underperformed on closed syntactic categories such as preposition, conjunction, and determiner.

The GGD is made up of 1,335 identifiers from 5 contexts: function, declaration, attribute, parameter, and class name. It represents a 95 and 6 sample from a dataset of identifiers from 20 open source systems \cite{newman_ensemble}. The CGD is made up of 1,275 identifiers, representing a 95 and 5 sample from a dataset of 30 systems validated similarly to prior work\cite{newman_ensemble}. The difference between these is that the CGD contains a higher population of closed-category words, such as prepositions, conjunctions, etc. We used srcML\cite{collard:2016} to do all data collection and filtering.

\begin{table}[]
\centering
\caption{Examples of grammar patterns}
\label{tab:closed-category-grammar-patterns}
\begin{tabular}{@{}ll@{}}
\toprule
Identifier Example & Grammar Pattern \\ \midrule
action to index map & N P NM N \\
as binary & P N \\
time for each line & N P DT N \\
server and port & N CJ N \\
open if empty & V CJ NM \\
adjust to camera & V P N \\ \bottomrule
\end{tabular}%
\end{table}

We combine these datasets to construct the Training Dataset (TD). This dataset contains a total of 1,335 + 1,275 = 2,610 identifiers. This translates to 7,173 rows, each row containing one word from the identifiers in our dataset. As stated, we train SCALAR using scikit-learn's GradientBoostingClassifier algorithm. The training is split set into train (70\%, 5021 words), and test (30\%, 2152 words) using a stratified, random sample. We stratified on the ground-truth PoS annotation for each word. We used stratified k-fold cross validation with k=10 for training to help improve the generality of SCALAR.

A number of features are used to help train the model. One of the major differences between the prior Ensemble Tagger \cite{newman_ensemble} and SCALAR is that it significantly reduces its reliance on external taggers, with NLTK\cite{bird2009natural} being the only other tagger whose output SCALAR is trained on. Instead, this tagger relies on word-embedding features and lexical features inspired by our prior work on grammar patterns \cite{Newman_GP, newman2021Catalogue}. This makes SCALAR much faster than its predecessor and equally, or more, accurate across the range of PoS categories. Due to space limitations, we will not go into detail about every new feature, but the strongest features in terms of their importance metric are as follows:
\begin{enumerate}
    \item \textbf{NLTK\_POS}. We use the NLTK part of speech tagger as a feature due to its speed, and it provides an off-the-shelf tagger perspective to SCALAR; it roots SCALAR in traditional PoS tagging, allowing it to focus on specializing the tags to the unique context of code.
    \item \textbf{Preposition Embeddings}. We collect a list of common prepositions (we reference word sources above the lists in the code\footnote{https://github.com/SCANL/scanl\_tagger/blob/master/feature\_generator.py}), translate them into word embeddings, then use an average, normalized vector of those embeddings to determine whether a given word is close (in terms of angles between vectors) to the general concept of a preposition. We do the same with \textbf{nouns} and \textbf{verbs} to create average noun/verb word embedding vectors.
    \item \textbf{Ratio of word position}. The position of a word within an identifier can provide valuable information about its role. For example, words at the end of an identifier tend to be nouns (i.e., head nouns); the word at the beginning of a function identifier tends to be a verb. This feature represents the ratio between given word's position and the length of the identifier that it is part of; it gives us an idea of how 'far' into an identifier a given word is.

\end{enumerate}


Our tagset is specialized for identifiers found in code. This tagset (shown in Table~\ref{tab:posusedtable}) is first discussed in Newman et al.'s original work on Grammar Patterns \cite{Newman_GP}. In this paper, Newman et al. shows that identifiers follow unique grammatical rules that are rarely in most natural human language text. Thus, they argue that specialized taggers are necessary for identifiers. Most of these tags are available and known to general PoS tagging approaches. However, there are two tags in our set that we must discuss, since their usage within identifiers is part of what sets the natural language in code apart from other natural language contexts, like newspapers. These are Noun Modifiers (NM)\cite{HillSWUM:2010, Gupta:2013, Newman_GP} and Preambles (PRE)\cite{HillSWUM:2010, Newman_GP}. A Preamble is an abbreviation which does one of the following: 
\begin{enumerate}
    \item Namespaces an identifier without augmenting the reader's understanding of its behavior (e.g., XML in XML\_Reader is not a preamble)
    \item Provides language-specific metadata about an identifier (e.g., identifies pointers or member variables)
    \item Highlights an identifier's type. When a preamble is highlighting an identifier's type, the type's inclusion must not add any new information to the identifier name.
\end{enumerate}

We give examples of each preamble type in the list. An example of (1) can be found in the GIMP and GLEW open-source projects, where GIMP and G\_ are namespace preambles to many variables. To discuss (2), we use Hungarian notation \cite{hungariannotation}. Hungarian notation is when developers, for example, put \textit{p\_} in front of pointer variables or m\_ in front of variables that are members of a class; any Hungarian notation in an identifier is considered a preamble. As an example of (3), given the declaration \textit{float* fPtr}, `f' in `fPtr' does not add any information about the identifier's role within the system, but reminds the developer that it has a type 'float'; \textbf{this is a preamble}. However, given an identifier \textit{char* sPtr}, `s' informs the developer that this is a c-string as opposed to a pointer to some other type of character array; `s' is \textbf{not} considered a preamble under this definition above. Intuitively, the reason for identifying preambles in an identifier is because they do not provide any information with respect to the identifier's role within the system's domain. Instead, they provide one of the types of information above. 

Another tag to note in Table \ref{tab:posusedtable} is \textit{noun modifier (NM)}, which is annotated on words that can be considered a pure adjective or noun-adjunct. A noun-adjunct is a word that is typically a noun but is being used as an adjective. An example of this is the word \textit{bit} in the identifier \textit{bitSet}. In this case, \textit{bit} is a noun which describes the type of \textit{set}, i.e., it is a set of bits. So we consider it a noun-adjunct. These are found in English (e.g., compound words), but generally not annotated as their own individual PoS tag. Prior work argues for the use of an individual tag for noun-adjuncts due to their ubiquity, and special role, in source code identifiers \cite{Newman_GP, HillSWUM:2010, Gupta:2013}.

Our evaluation is performed at the level of words and not full identifiers, since annotating even one word incorrectly within an identifier makes the annotation for the entire identifier incorrect. Word-level analysis is more granular and still correlates with higher correctness over whole identifiers.
\begin{table}[]
\centering
\caption{Test set metrics per tagger. Each tagger was run on the same test set, and metrics were gathered from their per-word performance.}
\label{tab:tagger_scores}
\resizebox{\columnwidth}{!}{%
\begin{tabular}{@{}lcccccc@{}}
\toprule
 &
  Accuracy &
  Balanced Accuracy &
  \begin{tabular}[c]{@{}c@{}}Weighted\\ Recall\end{tabular} &
  \begin{tabular}[c]{@{}c@{}}Weighted\\ Precision\end{tabular} &
  \begin{tabular}[c]{@{}c@{}}Weighted\\ F1\end{tabular} &
  \begin{tabular}[c]{@{}c@{}}Performance\\ (seconds)\end{tabular} \\ \midrule
SCALAR   & \textbf{0.8216} & \textbf{0.9160} & \textbf{0.8216} & \textbf{0.8245} & \textbf{0.8220} & \textbf{249.05}  \\
Ensemble & 0.7124 & 0.8311 & 0.7124 & 0.7597 & 0.7235 & 1149.44 \\
Flair    & 0.6087 & 0.7844 & 0.6087 & 0.7755 & 0.6497 & 807.03  \\ \bottomrule
\end{tabular}%
}
\end{table}

\begin{table*}[]
\centering
\caption{Category-level metrics for SCALAR based on test set performance}
\label{tab:category_level_metrics}
\resizebox{\textwidth}{!}{%
\begin{tabular}{l|ccc|ccc|ccc|ccc|ccc}
\hline
 &
  \multicolumn{3}{c|}{\begin{tabular}[c]{@{}c@{}}N \\ (Noun)\end{tabular}} &
  \multicolumn{3}{c|}{\begin{tabular}[c]{@{}c@{}}V \\ (Verb):\end{tabular}} &
  \multicolumn{3}{c|}{\begin{tabular}[c]{@{}c@{}}NM \\ (Noun Modifier):\end{tabular}} &
  \multicolumn{3}{c|}{\begin{tabular}[c]{@{}c@{}}D \\ (Digit):\end{tabular}} &
  \multicolumn{3}{c}{\begin{tabular}[c]{@{}c@{}}P \\ (Preposition):\end{tabular}} \\ \hline
 &
  \multicolumn{1}{l}{SCALAR} &
  \multicolumn{1}{l}{Ensemble} &
  \multicolumn{1}{l|}{Flair} &
  \multicolumn{1}{l}{SCALAR} &
  \multicolumn{1}{l}{Ensemble} &
  \multicolumn{1}{l|}{Flair} &
  \multicolumn{1}{l}{SCALAR} &
  \multicolumn{1}{l}{Ensemble} &
  \multicolumn{1}{l|}{Flair} &
  \multicolumn{1}{l}{SCALAR} &
  \multicolumn{1}{l}{Ensemble} &
  \multicolumn{1}{l|}{Flair} &
  \multicolumn{1}{l}{SCALAR} &
  \multicolumn{1}{l}{Ensemble} &
  \multicolumn{1}{l}{Flair} \\
Precision: &
  0.798 &
  0.7247 &
  \textbf{0.8854} &
  \textbf{0.7413} &
  0.5473 &
  0.5678 &
  0.8075 &
  \textbf{0.8687} &
  0.2516 &
  0.957 &
  0.9032 &
  \textbf{0.989} &
  \textbf{0.929} &
  0.6129 &
  0.8452 \\
Recall: &
  \textbf{0.8258} &
  0.8209 &
  0.5293 &
  \textbf{0.8514} &
  0.6322 &
  0.689 &
  \textbf{0.762} &
  0.6326 &
  0.6169 &
  \textbf{0.957} &
  0.9438 &
  0.8036 &
  \textbf{0.9351} &
  0.6934 &
  0.8506 \\
F1 Score: &
  \textbf{0.8116} &
  0.7698 &
  0.6625 &
  \textbf{0.7926} &
  0.5867 &
  0.6226 &
  \textbf{0.7841} &
  0.7321 &
  0.3574 &
  \textbf{0.957} &
  0.9231 &
  0.8867 &
  \textbf{0.932} &
  0.6507 &
  0.8479 \\ \hline
 &
  \multicolumn{3}{c|}{\begin{tabular}[c]{@{}c@{}}VM \\ (Verb Modifier):\end{tabular}} &
  \multicolumn{3}{c|}{\begin{tabular}[c]{@{}c@{}}PRE \\ (Preamble)\end{tabular}} &
  \multicolumn{3}{c|}{\begin{tabular}[c]{@{}c@{}}DT \\ (Determiner):\end{tabular}} &
  \multicolumn{3}{c|}{\begin{tabular}[c]{@{}c@{}}NPL \\ (Noun Plural):\end{tabular}} &
  \multicolumn{3}{c}{\begin{tabular}[c]{@{}c@{}}CJ\\ (Conjunction):\end{tabular}} \\ \hline
 &
  \multicolumn{1}{l}{SCALAR} &
  \multicolumn{1}{l}{Ensemble} &
  \multicolumn{1}{l|}{Flair} &
  \multicolumn{1}{l}{SCALAR} &
  \multicolumn{1}{l}{Ensemble} &
  \multicolumn{1}{l|}{Flair} &
  \multicolumn{1}{l}{SCALAR} &
  \multicolumn{1}{l}{Ensemble} &
  \multicolumn{1}{l|}{Flair} &
  \multicolumn{1}{l}{SCALAR} &
  \multicolumn{1}{l}{Ensemble} &
  \multicolumn{1}{l|}{Flair} &
  \multicolumn{1}{l}{SCALAR} &
  \multicolumn{1}{l}{Ensemble} &
  \multicolumn{1}{l}{Flair} \\
Precision: &
  0.75 &
  0.2917 &
  \textbf{0.9} &
  \textbf{0.701} &
  0.1856 &
  0 &
  \textbf{0.9697} &
  0.3434 &
  0.4444 &
  0.9204 &
  0.8761 &
  \textbf{0.9646} &
  \textbf{0.8235} &
  0.5294 &
  0.625 \\
Recall: &
  \textbf{0.8182} &
  0.3333 &
  0.2609 &
  \textbf{0.7816} &
  0.6 &
  0 &
  \textbf{0.8421} &
  0.7907 &
  0.7857 &
  \textbf{0.8525} &
  0.8319 &
  0.7899 &
  \textbf{0.9333} &
  0.5625 &
  0.8333 \\
F1 Score: &
  \textbf{0.7826} &
  0.3111 &
  0.4045 &
  \textbf{0.7391} &
  0.2835 &
  0 &
  \textbf{0.9014} &
  0.4789 &
  0.5677 &
  \textbf{0.8851} &
  0.8534 &
  0.8685 &
  \textbf{0.875} &
  0.5455 &
  0.7143 \\ \hline
\end{tabular}%
}
\end{table*}
\section{Evaluation}
We perform a comparison of our PoS tagger against an off-the-shelf part of speech tagger called Flair\cite{akbik2018coling}, as well as the previous iteration of our tagger, the Ensemble Tagger, which was shown to be the most accurate tagger compared to an off-the-shelf tagger (Stanford\cite{Toutanova:StanfordTagger}) and code-specialized taggers (SWUM\cite{HillSWUM:2010}, Posse\cite{Gupta:2013}) in prior work \cite{newman_ensemble}.

In order to compare with Flair\cite{akbik2018coling}, we need to translate Flair's output into our tagset; the translation between SCALAR's tagset and Flairs can be found in our Git repo README\footnote{https://github.com/SCANL/scanl\_tagger} where we show how to convert Penn Treebank to our tagset. In mapping to Penn Treebank, some granularity is lost. For example, we map most verb variation forms to just verb. This decision is based on prior experience with how verb variations are used in code, and is explained in more detail in prior work \cite{Newman_GP}. In summary, reducing these variations to verb simplifies the task of comparing, and many of these variations have uses in code that cause them to behave as non-verbs. For example, \textit{waitingList} has a verb (waiting), but it is being used as an noun modifier (describing the type of list). Refer to \cite{Newman_GP} for more on that issue. Note that the purpose of comparing to Flair is primarily to show that an off-the-shelf PoS tagger cannot be readily used to annotate with the specialized grammatical structure of identifiers. It is not designed to recognize this specialized structure and, as a result, it significantly under-performs versus its relatively high accuracy on normal PoS tagging tasks. That is, even if we use Flair's tagset, it will still under-perform. This is clear from the performance on the NM category in Table \ref{tab:category_level_metrics}. Instead of recognizing identifiers like bitSet as a noun-adjunct and a noun (NM N), Flair recognizes it as two nouns (N N). This does not correctly identify the relationship between these words. For more information, and examples, refer to Newman et al.\cite{Newman_GP, newman2021Catalogue}. 

Compared with the Ensemble Tagger\cite{newman_ensemble}, its predecessor, SCALAR is somewhat better in terms of accuracy, precision, recall, and F1, but the true advantage that SCALAR has above its predecessor is \textit{speed}. SCALAR is much faster, annotating all 2152 words in the test data set in 249.05 seconds, versus the 1149.44 seconds it took the Ensemble Tagger. That's 0.12 (249.05/2152) seconds per word versus 0.53 (1149.44/2152) seconds per word; SCALAR is 4.42 times faster while remaining more effective than its predecessor.

In terms of overall performance, SCALAR generally outperforms the other two taggers on identifiers in terms of our performance metrics at a macroscopic level (Table \ref{tab:tagger_scores}), and on a per-category basis (Table \ref{tab:category_level_metrics}).


\section{The Application}

SCALAR runs as a Python Flask server using Waitress, which opens the tagger to a user-defined address and port and has the ability to handle simultaneous web requests from users. This allows SCALAR to be available for an internal group to use. HTTP requests are sent to SCALAR that contain the identifier to process. SCALAR is very customizable, and allows for user specified acceptable words and abbreviations. Thus any domain-specific terms that do not have a standard dictionary definition are manually flagged as valid words and reported with corrected PoS information.

SCALAR returns JSON output with a list containing each word found in an identifier. For each word, the output includes PoS information and an additional tag indicating if it is a standard dictionary word.  Every time SCALAR encounters an identifier for the first time, it caches the results of the splitter and tagger. Every subsequent time SCALAR encounters the same identifier, it returns the cached information on the identifier. This caching considerably speeds up the splitting and tagging process, eliminating the need to process an identifier which has been previously tagged. This provides a significant performance increase for users who work in code bases where the same identifiers are reused frequently. Running the application on a couple systems we found that the average time for a result of an identifier for the first time is 133.2 milliseconds and after caching (second time) the average time for a result for the same identifier is 1.2 milliseconds.  

Additionally, SCALAR saves the first time it encounters an identifier, the most recent encounter, and the number of times it encounters an identifier. This provides a log of identifiers and is useful for researchers to better understand identifier uses. The first and last encounters are saved as UNIX timestamps and displayed in the JSON output.


SCALAR is available for download as a Docker image. The Docker image automates the process of setting up an environment in which SCALAR can run by including all of the dependencies in a container. The Dockerfile included with SCALAR downloads the required Python packages, word embeddings, English word dictionary, and a list of allowable domain-specific words and abbreviations. Downloading this information from a separate source every time the docker image is built allows the lists to be easily updated by rebuilding the Docker image. Once these items are downloaded, the docker image automatically runs the commands to train the tagger and start the Python Flask server to receive requests. The port on which SCALAR listens for requests is mapped to port 8080 (a standard port for web traffic) on the machine hosting the docker container. As requests are sent to SCALAR, a JSON file containing the result cache is stored in a docker volume. Storing the cached output allows for data to be collected about the identifiers sent to the tagger during a specific time frame. 

\section{Conclusion}
This paper introduces SCALAR, a part-of-speech tagger for source code identifiers. It is trained using several features involving word embeddings and an external part-of-speech tagger output. It is designed to support the generation of grammar patterns \cite{Newman_GP, newman2021Catalogue, peruma_test_name_gp} for the purpose of analyzing, critiquing, and improving identifier names. SCALAR is faster than its predecessor, has similar performance, and out-performs other PoS taggers on identifier names. 
\newpage
\bibliographystyle{IEEEtran}
\bibliography{references}
\end{document}